# Hybrid DEEC: Towards Efficient Energy Utilization in Wireless Sensor Networks

M. Y. Khan[1], N. Javaid[1], M. A. Khan[1], A. Javaid[2], Z. A. Khan[3], U. Qasim[4]

[1]COMSATS Institute of Information Technology, Islamabad, Pakistan.

[2]Mathematics Dept., COMSATS Institute of Information Technology, Wah Cantt., Pakistan.

[3]Faculty of Engineering, Dalhosie University, Halifax, Canada.

[4]University of Alberta, Alberta, Canada.

*Abstract*— **The clustering algorithm are considered as a kind of key technique used to reduce energy consumption. It can help in increasing the stability period and network life time. Routing protocol for efficient energy utilization should be designed for heterogeneous Wireless Sensor Networks (WSNs). We purpose Hybrid-DEEC (H-DEEC), a chain and cluster based (hybrid) distributed scheme for efficient energy utilization in WSNs. In H-DEEC,elected Cluster Heads (CHs) communicate the Base Station (BS) through beta elected nodes, by using multi-hopping. We logically divide the network into two parts, on the basis of the residual energy of nodes. The normal nodes with high initial and residual energy will behighlyprobable to be CHs than the nodes with lesser energy. To overcome the deficiencies of H-DEEC, we propose Multi-Edged Hybrid-DEEC (MH-DEEC). In MH-DEEC the criteria of chain construction is modified. Finally, the comparison in simulation results with other heterogeneous protocols show that, MH-DEEC and H-DEEC achieves longer stability time and network life time due to efficient energy utilization.**

# I.    INTRODUCTION

Recent technological advancements in hardware have made it possible to have extremely small and low powered devices, equipped with programmable computing, multiple parameter sensing and wireless communication capability. Due to low cost, micro-sensors are preferred for monitoring the physical environment in terms of sensing temperature, moisture, light, sound, motion, pressure etc. These sensor nodes collect data from sensing field and after aggregating send the gathered information to end user. Since WSNs are usually exposed to unpredictable and dynamic environments, it is possible that nodes might lose their connectivity. Conventional centralized routing schemesshould work with global knowledge of the entire network, and a transmission failure of a critical node will potentially cause a serious problem for the network [1].

Clustering can be done in two types of networks, homogeneous and heterogeneous networks. In homogeneous networks, initial energy of all nodes is same while in heterogeneous network, all nodes have different initial energy. In [2], a comprehensive comparison of different routing techniques, including clustered and non-clustered techniques is done, which shows clear picture of performance comparison between them.

In this paper, we study the performance of the clustering algorithms for saving energy inheterogeneous sensors field. Usually clustering schemes LEACH [3] and DEEC [4] are described as, each node transmits sensed data to the BS, through a CH. CHs are elected periodically by certain criteria, send aggregated data of cluster members to BS, from where the end user can access the data. In previous schemes [3] [4] [5], it was unfair with CHs to directly communicate to BS, It imposes extra burden on the nodes, that are elected as CHs. We assume that all nodes of the network are with different initial energy, which is source of heterogeneity. H-DEEC and MH-DEEC permit the network for balanced energy utilization. High energy nodes of the network are responsible for transferring the data (received from CHs) to BS.

# II.    RELATED WORK AND MOTIVATION

There are two kinds of clustering schemes. The schemes applied in for the networks having similar nodes initial energy is referred as homogenous clustering schemes and the algorithm applied in an environment where initial energies of nodes are different from each other are called as heterogeneous clustering schemes. It is difficult to implement an energy awareheterogeneous clustering algorithm due to the complex energy configuration of the network. Some of the previous clustering algorithm for homogenous networks are LEACH, PEGASIS [6] and HEED [7] and non-homogenous one's are SEP and DEEC.

A number of schemes were proposed to make the communication more energy efficient. Like LEACH was devised for clustering in homogeneous networks. LEACH selects CH periodically and drains energy uniformly. Each node decides it self, whether or not to be a CH. It performance badly degrades in heterogeneous networks as shown in [5]. In PEGASIS nodes are supposed to form a chain, which can be computed by neighbour node or the BS. The requirement of Global Positioning System (GPS) for every node make this method difficult to implement and not viable for localizing sensor nodes [8]. SEP and CEEC [9] are also heterogeneous routing approaches, but these protocols are not considered as realistic approaches due to their limited levels of heterogeneity. HEED is a distributed clustering algorithm, which selects the CH stochastically. But HEED is not suitable for heterogeneous environment due to its CH selection criteria. DEEC is an energy efficient approach for heterogeneous networks, which selects the CHs on the basis of initial and residual energy.

In [10] a detail survey and performance comparison of classical DEEC with its flavours is done. To avoid complexity in our proposed scenarios, we use DEEC routing scheme for clustering propose.In order to reduce the energy consumption, sometime chain based routing schemes are used for WSNs.

In order to reduce the energy consumption, sometime chain based routing schemes are also used for WSNs. There are many chain-based protocols other than PEGASIS, like EEPB [11] IEEPB [12] and EECB [13]. Aiming to solve the problem of Long Link (LL) FengSenet. al. [12] proposed IEEPB which avoids the formation of LL, but still it cannot utilize the energy efficiently. There are number of chain based protocol, which uses different scenarios of chain construction, other than on the basis of nearest neighbour node like [14].

There are several classical approaches with large network life time. Weanalyse that, for filling the coverage holes in the network, there should be a protocol which utilizes energy efficiently to increase the stability time. By increasing stability time, maximum coverage can be provided to the network. Our proposed schemes H-DEEC and MH-DEEC fulfil these requirements. Our proposed scenarios are based on DEEC, with changing the cluster-head to BS communication method. Simulation Results shows that H-DEEC and MH-DEEC performs better than DEEC and SEP, which are classical heterogeneous routing approaches.

## III. PROPOSED H-DEEC

H-DEEC is based on DEEC scheme for heterogeneous networks, where all nodes use initial and residual energy level for cluster-head election. Each node in the network has the information of all the other fellow nodes. In this section, we consider a N nodes network, randomly deployed over a region of M×M. We assume that BS is located outside the network. Figure2 shows the network model of H-DEEC with heterogeneous environment.

In order to achieve an acceptable Signal to Noise Ratio (SNR) in transmitting a L-bits message over a distance $d$, we use radio model similar to that one in [15].

***Hybrid distributed energy efficient clustering***

To increase the stability time and for efficient energy utilization, we propose H-DEEC routing scheme for WSNs. In our proposed scenario, we logically divide the network into two parts, on the basis of residual energies of the nodes; normal nodes and beta nodes.

In this heterogeneous environment, 90% of nodes act as normal nodes and reaming 10% are beta nodes. Normal nodes make clusters and send their information to CHs. Beta nodes will collect the aggregated data from CHs as shown in Figure 2. Beta nodes are elected on the basis of energy, mean that 10% nodes with highest energy are the beta nodes. Residual energy of beta nodes is always greater than normal nodes in ourproposed scenario. To make energy utilization efficient, beta node election process will be made in every round. The H-DEEC scheme is combination of two scenarios; clustering and chain construction.

## Figure 1

**1)** ***Clustering***: As mentioned above, part of the network comprises with normal node will follow the clustering scenario as done in DEEC and as shown in Figure 2. All nodes in the network are aware of their fellow nodes energy level and position. Normal node will follow the same strategy in terms of estimating average energy of the network, and CH selection algorithm, probability of each node to become CH is based on residual energy. For network of N nodes and an additional energy factor, it can be calculated as:

$$p_i = \frac{p_{opt}N(1+a)E_i(r)}{(N+\sum_{i=1}^{N}a_i)\bar{E}(r)} \tag{1}$$

Where $p_{opt}$ is the reference value of the average probability$p_i$, which determine the rotating$e_{poc}$.$\bar{E}(r)$is average energy of $r_{th}$ round and$E_i$ is the residual energy of node $s_i$ at round $r$. We can estimate average energy at $r_{th}$ round as follow:

$$\bar{E}(r) = \frac{1}{N}E_{total}\left(1-\frac{r}{R}\right) \tag{2}$$

$R$denotes the number of rounds a network will be alive and is estimated as:

$$R = \frac{E_{total}}{E_{round}} \tag{3}$$

**2)** ***Chian Construction:***Chain formationof beta nodes in our proposed scenario is done on the basis of a classical chain-based routing scheme PEGASIS. It saves a significant amount of energy as compared with the other routing protocols, due to its improvement in delivery of data. Beta nodes will be connected by following greedy algorithm. BS initiates the chain forming process by marking the farthest node. Farthest node finds its nearest neighbour and so on. Leader node of the chain is selected as the beta node with least distance from BS. As chain rotates in every round on the basis of energy and beta node chain always will be diverted towards the BS.

In DEEC CHs were supposed to do communication directly to BS, which consumes a large amount of energy. In our proposed scenario, CHs of the normal nodes send their aggregated data to the nearest beta node. After aggregating the data received from previous chain members, it sends it to BS (located far away from network). In other words beta nodes transmit the received data hop by hop to BS. Due to larger consumption of energy in direct communication, stability time of network decreases abruptly. By such type of communication coverage holes (no coverage areas) are created in the network. In H-DEEC, CH communicates with BS by the mean of beta nodes, which follow multi hopping scheme. This approach will distribute load evenly among the sensor nodes, due to which unstability time decreases.

## IV.    PROPOSED MH-DEEC

After making some modifications in H-DEEC, we propose another scheme MH-DEEC. In this protocol, chain forming scenario is a bit change. In MH-DEEC node distribution ratio is same as in H-DEEC with similar radio model and clustering scenario. Like H-DEEC, chain construction process is initiated by BS by marking the farthest beta node from BS. In this scenario every node connects itself to its nearest neighbour node, doesn't matter whether that node is already connected or not. This scenario leads to formation of a Multi-Edged Chain as done in IEEPB. MH-DEEC chooses the chain leader of beta nodes using weighting method, which requires both residual energy and distance of each beta node from the BS.

Distance parameter $D_{toBS}$ is formulated by multi-path model given as:

$$D_{toBS} = \frac{d_{toBS}^4}{d_{avg}^4} \qquad (4)$$

Where, $d_{toBS}$ is the distance of beta node from BS and $d_{avg}$ represents the average distance between beta nodes and BS.

Energy parameter $E_p$ is calculated as follows:

$$E_p = \frac{E_{init\_b}}{E_{i\_b}\ (r)} \qquad (5)$$

Where, $E_{init\_b}$ is the initial energy of the beta node and $E_{i\_b(r)}$ is the residual energy.

Combined weight is calculated as:

$$W_i = w_i E_p + w_2 D_{toBS} \qquad (6)$$

Where $w_1$ and $w_2$ are weight factor meets the formula of:

$$w_1 + w_2 = 1 \qquad (7)$$

For leader selection, weights are compared; the beta node with minimum weight will be selected as leader of beta nodes for that round. Leader node is responsible for sending the data, received from other beta nodes and CHs to the BS. Chain forming procedure followed in our proposed MH-DEEC also solves the problem of Long Link, which reduces the energy efficiency of nodes. Node connectsitself to its nearest neighbour node which forms a multi-edge chain network as shown in Figure 3.

<p style="text-align:center; color:red;">Figure 2</p>

Algorithm 1 defines, how nodes will communicate with BS through CHs and beta nodes in a *N* nodes network. *r* is the number of rounds which indicates the time. As mentioned that first normal nodes of $i_{th}$ cluster will send their data *data_Normal_i* to respective cluster-heads. Cluster-heads will pass their received data *(Rec_data_CH_i)*to nearest beta node. Finally beta nodes will be responsible for delivering the gathered data *(Rec_data_Beta_j)* to the BS, through multi-hopping.

According to radio model followed in this case, energy consumption is directly proportional to distance between transmitter and receiver. By following the chain formation process of IEEPB, distance to BS is decreased, due to which a significant amount of energy is saved by the beta nodes.

<p style="text-align:center; color:red;">**Algorithm 1**</p>

<p style="text-align:center;">V.     SIMULATIONSANDRESULTS</p>

In this section, we evaluate the performance of H-DEEC and MH-DEEC using MATLAB. We consider a 100*m*x100*m*, randomly deployed, heterogeneous network of 100 nodes. Like DEEC, we ignore the effect of signal collision and interference in wireless channel. It is assumed that the BS is fixed and located far away from the network at (30,150). The simulation parameters are given in Table 1.

<p style="text-align:center; color:red;">Table 1</p>

The protocols compared with H-DEEC and MH-DEEC, include DEEC and SEP. We executed simulations for 4000 rounds. For obtaining the comparison results, we have done independent simulations (i-e starting from different random number seeds). Figure 4 represents the number of alive nodes during the network lifetime. Comparison is done in term of the live nodes over rounds. From figure we observe that stability time of the H-DEEC is far better than that of SEP and DEEC, which is approximately 1700 rounds. After modifying the chain construction and leader selection criteria in MH-DEEC, it performs better even than H-DEEC. From results comparison, we observe that unstability time of H-DEEC is less than that of MH-DEEC which is positive feature of H-DEEC. Network life time of MH-DEEC is still better than other protocols. SEP has a larger network lifetime than DEEC and H-DEEC and MH-DEEC, because in SEP, advanced nodes die more slowly than the normal nodes. The slow death rate of advanced nodes in SEP causes larger unstability period (until the last node die). H-DEEC and MH-DEEC have shorter unstability periods, which are around 716 and 1156 respectively, as shown in Figure 4. In our proposed schemes unstability time is less because energy distribution is done efficiently. As in both H-DEEC and MH-DEEC beta node selection scenario is same, top ten energy nodes (beta nodes) of the network are selected for chain forming. Simulation results clearly shows that by introducing beta nodes in every round increases the stability period and decreases the unstable time. By changing the leader selection method in MH-DEEC, network survives for greater time. Efficient energy utilization become possible due to rotation of chain making process in every round (i-e.chain is not fixed).

## Figure3

In SEP and DEEC, first node dies at 922 and 965 respectively while in H-DEEC and MH-DEEC, first node dies at 1487 and 2389 respectively. So, stability time of H-DEEC and MH-DEEC is 35% and 60% better than DEEC respectively and 38% and 61% better than SEP due to inefficient energy utilization in these classical protocols. In DEEC the whole network dies at 2401, while in MH-DEEC network lifetime is 3545 rounds, so network life time in MH-DEEC is 32% less in than that of the conventional DEEC. By rotation of the chain forming process in our proposed schemes, energy is utilized efficiently.

Figure 5 shows the comparison in term of number of packets received at BS over rounds. The simulation results show that the throughput for SEP and DEEC is extremely low as compared to MH-DEEC and H-DEEC. As in Figure 4, it is shown that the network lifetime of SEP and DEEC was significantly greater than H-DEEC, but this does not mean that the nodes will transmit more packets to BS (i-e the throughput is low), even worse during last rounds, when live nodes density is significantly low. In case of MH-DEEC unstable time is less than classical heterogeneous protocols but H-DEEC still out performs in terms of unstability time. In MH-DEEC, this drawback is overcome by achieving larger network life time, which is about 3545 rounds. During unstable time, a-lot of empty spaces (in term of the coverage) are created, due to which network gets sparse and these spaces get more and more wider. As shown in Figure 4 after 50% nodes die, it takes less than 10 rounds for other 50%. Due to short unstability period, H-DEEC and MH-DEEC provide better coverage than other two protocols,because of this

## Figure4

BS will receive more packets in the proposed scenarios. From simulation results in Figure 5, it is observed that the throughput of MH-DEEC is 57% and 93% better than DEEC and SEP respectively and throughput performance of H-DEEC is 45% and 90% better respectively. Significant difference in throughput of our proposed schemes with DEEC and SEP is because of providing full coverage for the most of the time to network (i-e. stability period is larger).

## Figure 5

For simulations, nodes begin with limited initial energy. Once a node runs out of energy, it is considered as dead and no longer transmits or receives data. Figure 6 presents the comparison of energy consumption of the H-DEEC and MH-DEEC with other two protocols. Initial energy of nodes is random, starting with different random number seeds. Due to independent simulations total initial energy of network in all cases (MH-DEEC, H-DEEC, DEEC and SEP) is different. Simulation results show the residual energy of network over rounds. Random distribution of energy is done to meet the real time scenario. In Figure 6, the graph indicates the rate of energy consumption. Higher the slop, faster the energy consumption will be. In MH-DEEC energy is utilized efficiently while H-DEEC consumes energy somehow better than classical routing scenarios. In DEEC,CHs communicate directly with BS, which consumes a lot of energy. As in the proposed scenarios, CHs send their aggregated data to BS.This process saves a large amount of energy. Beta nodes are elected in every round on the basis of residual energy, communicate through multi-hoping, this leads to efficient energy utilization.

From simulation results, it is observed that the rotated beta nodes chain making scenario in every round, helps H-DEEC to perform better than DEEC and SEP in terms of stability time, throughput and energy utilization. However, with modifications in chain construction scenario in H-DEEC, our second proposed scheme MH-DEEC outperforms other schemes including H-DEEC. Multi-edge chain making and leader selection criteria in MH-DEEC make it even better than H-DEEC.

## VI. CONCLUSION

H-DEEC and MH-DEEC routing protocol are proposed as energy aware adaptive clustering protocols for heterogeneous WSNs. In H-DEEC the network is divided into two parts on the basis of initial and residual energy. Normal nodes will elect themselves as a CH and Beta nodes will do collect data from CHs and send it to BS by multi-hopping. MH-DEEC is proposed by making modifications in chain making and leader selection scenario in H-DEEC, it performs even better than H-DEEC. Unlike SEP and DEEC, H-DEEC and MH-DEEC can perform well in a heterogeneous wireless sensor field. Moreover, it also considers the problem of locating BS outside the network. MH-DEEC and H-DEEC look promising; there are still many challenges like lesser unstability time, sensor nodes localization and interference among the sensor nodes that need to be solved in routing of sensor networks. In future, many issues like sink mobility, beta node selection scenario and CH election criteria are to be discussed.

# Tables in the paper

Table 1

Parameters used in Simulations

| Parameter | Value |
|---|---|
| Network size | $100m \times 100m$ |
| Number of nodes | 100 |
| BS position | $(30m, 150m)$ |
| Packet size | 4000 bits |
| $P_{opt}$ | 0.1 |
| $E_0$ | 0.5 J |
| $E_{elec}$ | 5 nJ/bit |
| Distance threshold $(d_0)$ | $70m$ |

---

**Algorithm 1 MH-DEEC Algorithm**

1.  $r \leftarrow Number\ of\ rounds$
2.  $N \leftarrow Number\ of\ nodes\ in\ Network$
3.  $Beta\_Node \leftarrow j_{th}\ Beta\ elected\ node\ of\ the\ chain$
4.  $Normal\_Nodes_i \leftarrow Normal\ nodes\ associated\ to\ i_{th}\ cluster$
5.  $CH_i \leftarrow Cluster\ Head\ of\ the\ i_{th}\ Cluster$
6.  $Rec\_data\_CH_i \leftarrow Recieved\ data\ by\ i_{th}\ Cluster\ head$
7.  $Rec\_data\_Beta_i \leftarrow Recieved\ data\ by\ nearest\ Beta\ Node\ to\ i_{th}\ cluster$
8.  $Data\_Normal_i \leftarrow Data\ of\ Normal\ Nodes$
9.  $Leader \leftarrow Index\ of\ leader\ node\ of\ the\ round\ r$
10. **if** $node == Normal\_Node$**then**
11.     **if** $Normal\_Node \neq CH$**then**
12. $Rec\_data\_CH_i == Data\_Normal_i$
13.     **else**
14.     **If** $Normal\_Node == CH$ **then**
15. $Rec\_data\_CH_i == Rec\_data\_CH_i$
16.     **end if**
17.         **end if**
18.     **end if**
19.     **while** $j \neq Leader$**do**
20. $Rec\_data\_Beta_{j+1} = Rec\_data\_beta_{j+1} + Rec\_data\_Beta_j$
21.     **end while**

# FIGURES

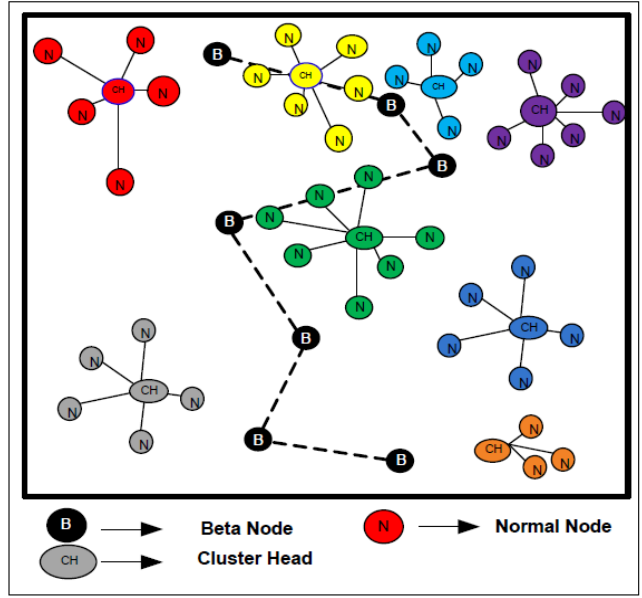

Figure 1. Network model of H-DEEC

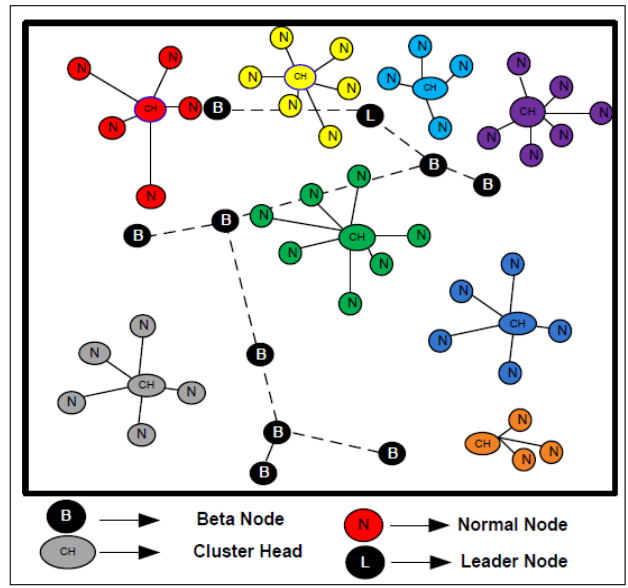

Figure 2. Network model of MH-DEEC

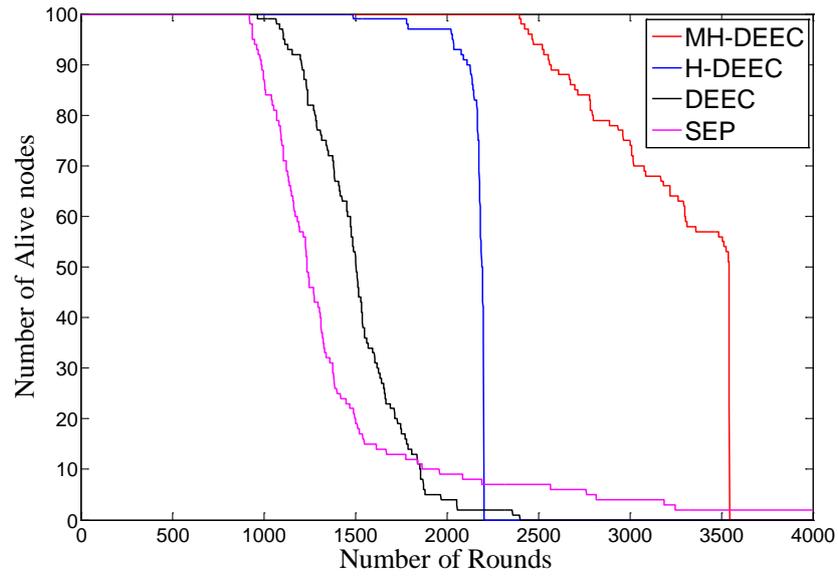

Figure 3. Performance consumption on the basis of stability time

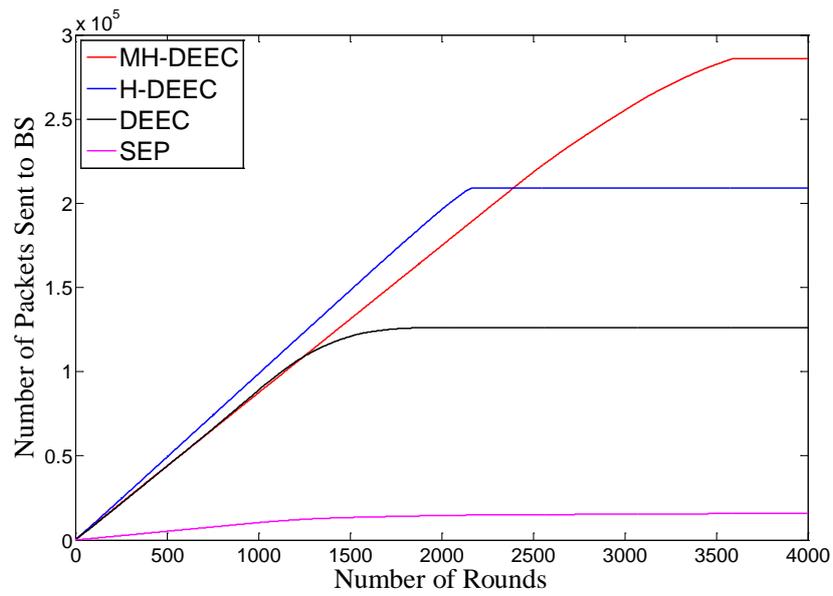

Figure 4.  Comparison of MH-DEEC, H-DEEC, DEEC and SEP in throughput

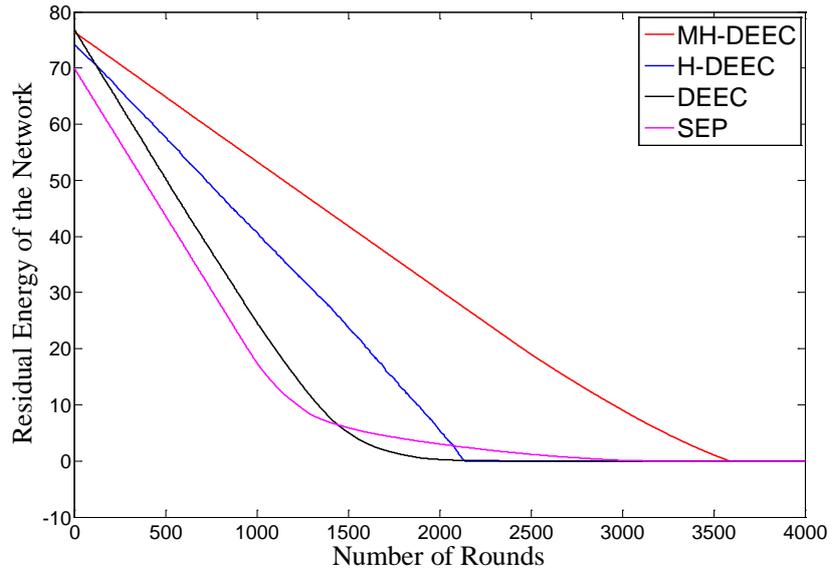

Figure 5. Residual energy of 100×100 network over rounds